\documentstyle[sprocl,epsfig]{article}

\def\PLB{{\em Phys. Lett.}  B}
\def\PRL{\em Phys. Rev. Lett.}

\def\be{\begin{equation}}
\def\ee{\end{equation}}
\def\bea{\begin{eqnarray}}
\def\eea{\end{eqnarray}}

\begin{document}
\title{UNCERTAINTY OF PREDICTED HIGH Q\( ^{2} \) STRUCTURE 
FUNCTIONS DUE TO PARAMETRIZATION ASSUMPTIONS}

\author{A. Caldwell, S. Paganis, F. Sciulli}

\address{Columbia University, Nevis Laboratories, 
P.O.Box 137, Irvington,\\ NY 10533, USA\\
E-mail: paganis@nevis1.columbia.edu}


\maketitle\abstracts{
The HERA luminosity upgrade
is expected to provide statistically significant measurements of the proton
structure functions at 0.5\textless{}$x$\textless{}0.7\(  \) 
and very high Q\( ^{2
} \)(Q\( ^{2}>>M_{Z}^{2} \)).
The behaviour of the parton densities (PDFs) in this high $x$, 
Q\( ^{2} \) regime
is predicted from DGLAP evolution of 
PDF parametrizations from lower Q\( ^{2} \)
fits of the data. Uncertainties in the PDFs at high $x$ may propagate to lower
$x$ through DGLAP evolution at very high Q\( ^{2} \). In this presentation the
behaviour of the PDFs at high $x$ is reexamined.
We present 
the effects at Q\( ^{2}=40000GeV^{2} \)
and compare our results with uncertainties 
obtained from propagation of experimental
errors at high $x$, high Q\( ^{2} \) DIS data.
}

\section{Introduction}
The HERA and Tevatron upgrades will allow experiments to test the 
behaviour of PDFs at very high $x$ and $Q^2$. Violations of the 
predicted PDF behaviour may signal new physics beyond the 
standard model. Thus, it is 
important to understand to what extent the predicted PDF behaviour 
depends on the parametrization assumptions. Currently there is no 
systematic study that estimates uncertainties due to parametrization 
assumptions. 

\noindent
In the past, 
a number of attempts to 
calculate PDF errors at high $Q^2$ 
was made.
Recently \cite{botje2}, a large set of DIS data was fit
and the error in DIS 
differential cross section was estimated 
as a function of $x$, $Q^2$ and $y$,
taking into account all systematic and statistical errors. 
A conventional PDF parametrization was used:
$xq(x)=N_{q}x^{A_{q}}(1-x)^{B_{q}}(1+C_{q}x)$.
The quoted error 
on the neutral current (NC) DIS  cross section
$d\sigma^{NC}/dx$ 
at $x=0.75$ is
$\simeq 0.1$.
Yang and Bodek \cite{ybod} introduced a modification 
to the $d/u$ ratio 
$(\frac{d}{u})^\prime = \frac{d}{u} + 0.1x(x+1)$
which lead to better agreement
with recent charged current 
data from HERA. 
Finally, 
Kuhlmann et al
\cite{cteq} added an extra term to the 
conventional up valence quark parametrization 
in order to investigate if such modifications could explain 
the reported HERA excess at high $Q^2$ in 1997. 
Such a term generates a significant excess for $x\geq 0.75$:
$xu(x)=xu(x)_{CTEQ4}^{}+0.02(1-x)^{0.1}$.
Their study showed that even large modifications at high $x$ 
could not explain the HERA excesses. The quoted modification 
in $F_2$ at $x=0.75$ and $Q^{2}=40000GeV^{2}$ is:
$\Delta F_2/F_2^{NC}<0.3$.
Although this study for the first time questioned the induced 
error in $F_2$ due to modifications to the PDFs, it did not 
cover the full parameter space of possible modifications. 
Also, the extra term $0.02(1-x)^{0.1}$
is problematic as $x\rightarrow 0$ when the valence sum rule is 
calculated.

\section{Systematic Modification of the PDFs at high x}
\begin{figure}[hbtp]
\parbox[t]{2.4in} { 
\psfig{figure=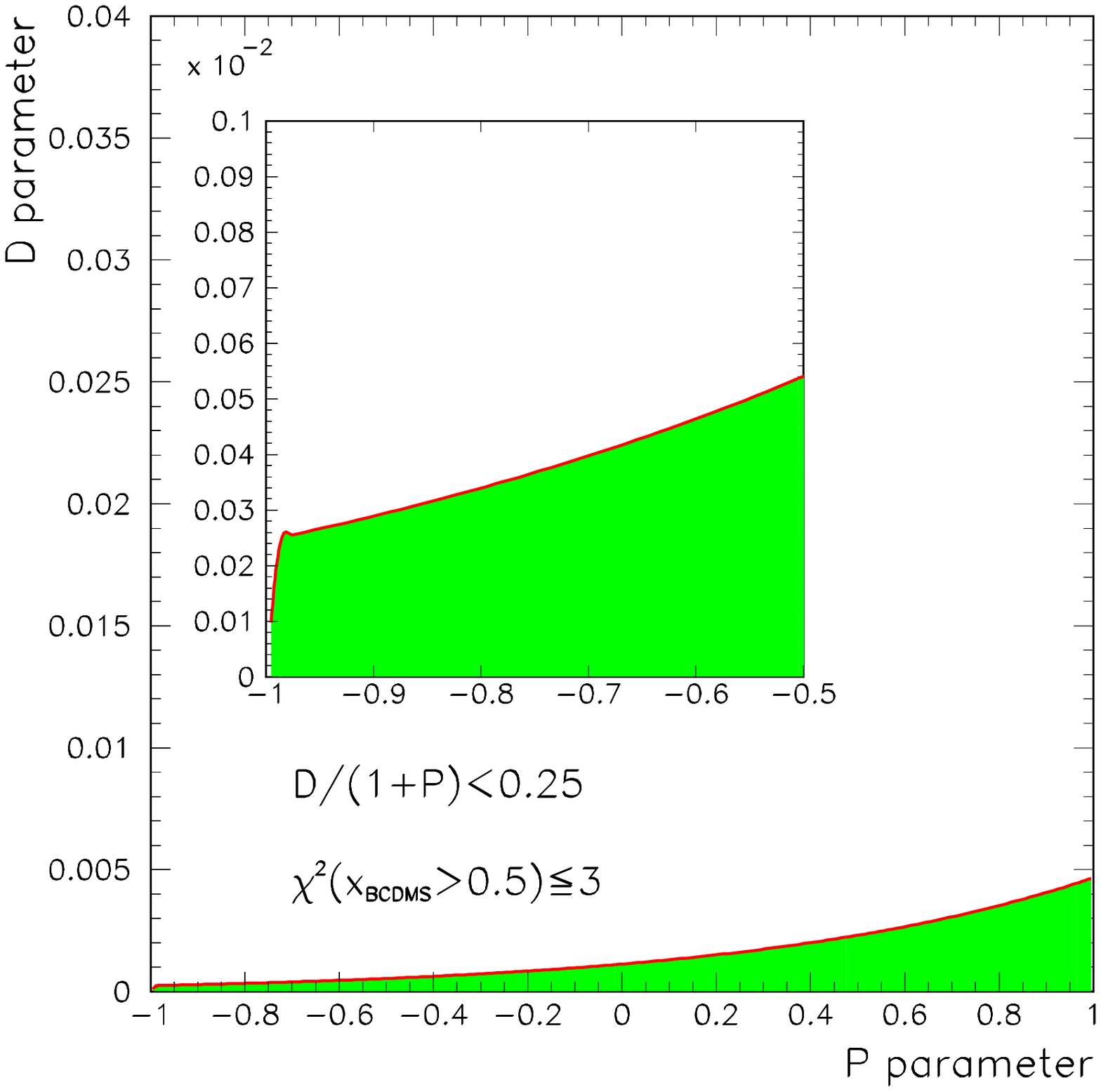,width=2.4in}
\caption{Allowed P,D parameter space.
\label{pspace}
}}\ \hspace{-0.2in} \
\parbox[t]{2.4in} {
\psfig{figure=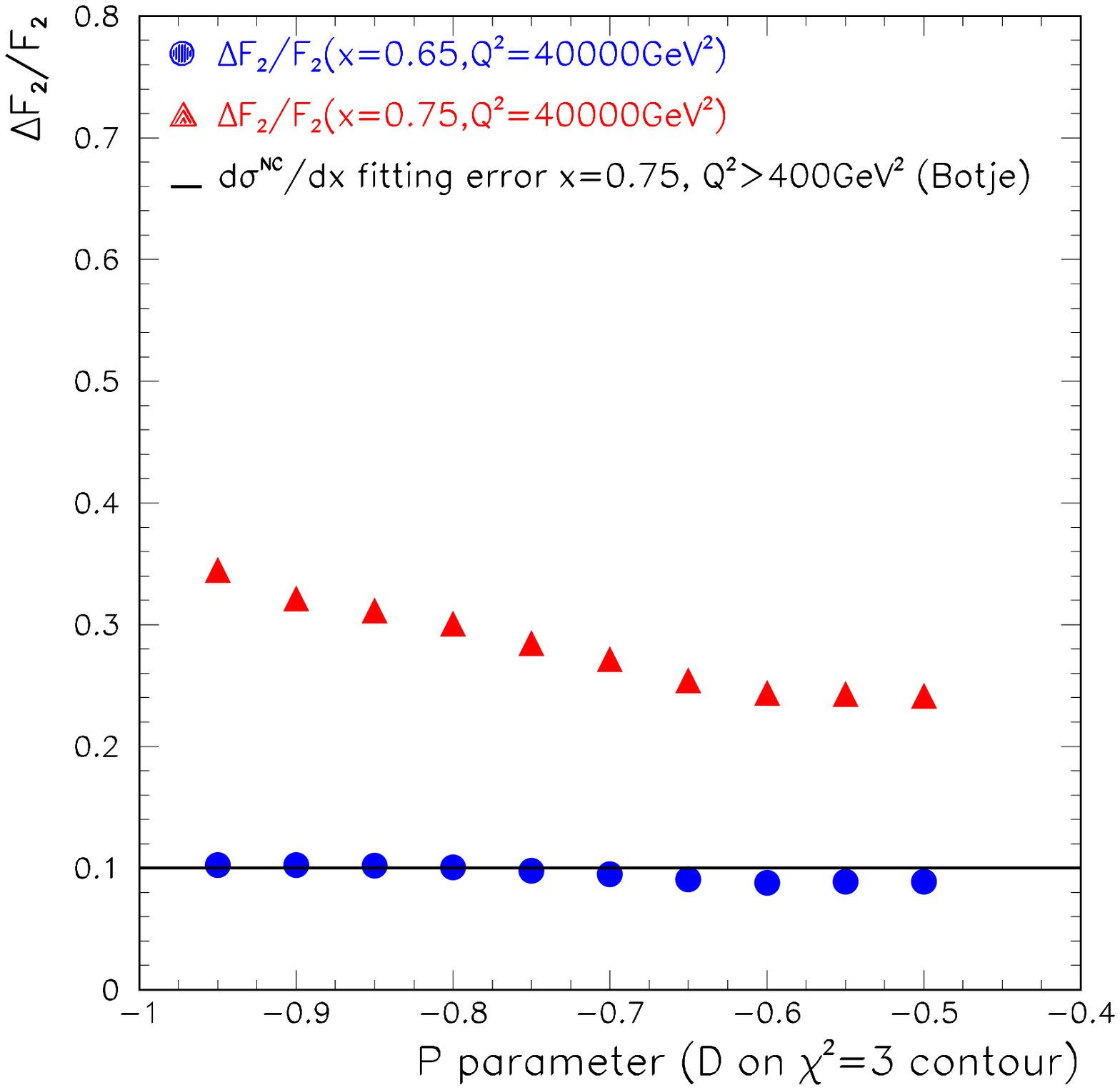,width=2.4in}
\caption{Modifications in $F_2$ for D and P values at the 
boundary of the parameter space.
\label{results}
}}
\end{figure}

\begin{figure}[hbtp]
\begin{center}
\mbox{\epsfig{figure=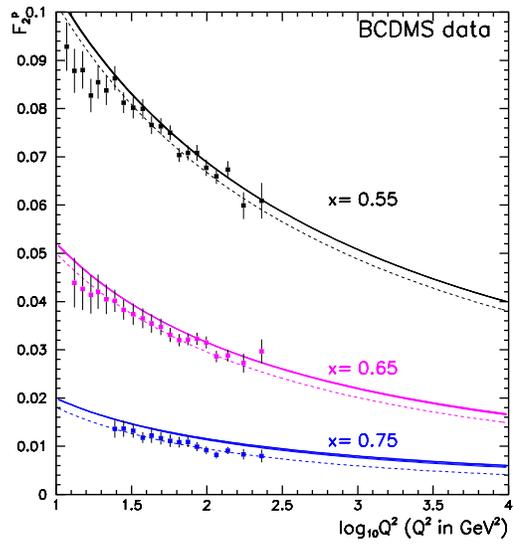,width=3.0in}}
\caption{Comparison between modified $F_2$ distributions and
BCDMS data.}
\label{fqual2}
\end{center}
\end{figure}

In this section a new approach is presented. It was found 
that the simplest two-parameter modification (addition in our case)
in the standard up quark density which produces sum rules free 
from infinities, is:
\begin{equation}
xu(x)=xu(x)_{CTEQ4}+Dx(1-x)^{P}
\label{columbia}
\end{equation}
The extra term $Dx(1-x)^{P}$ diverges as $x\rightarrow 1$ for 
negative powers P. This extreme behaviour 
of the PDF can be regulated by adjusting the strength parameter 
P, so that it does not violate any experimental observations.
In this study the Yang-Bodek correction is included:
\( xd(x)^{\prime }=xd(x)_{CTEQ4} \)\( +0.1x(x+1)xu(x) \).
We fit DIS fixed target data for \( x>0.01 \)
and \( W^{2}>10GeV^{2} \) (Datasets: NMC, BCDMS, SLAC, E665).
QCDNUM \cite{botje2} was used to evolve parametrizations.
The available value space for the extra term is constrained 
by experimental data. Our goal is to find the region of the D,P 
parameter space that maximizes dF\( _{2} \)/F\( _{2} \)
at high Q\( ^{2} \). We constrain the D and P parameters:
$\Delta (xu_v(x)+xd_v(x))/(xu+xd) \leq 0.2$ (Total momentum constraint),
$\chi^2(x_{BCDMS}>0.5)/dof \leq 3$.
In figure \ref{pspace} the allowed parameter space is shown. In figure 
\ref{results} we pick up pairs of (P,D) points from the boundary of the 
param. space and calculate the effects of the corresponding 
extra term at very high $Q^2$. As shown, even extreme
modifications of the structure functions at high $x$ cannot produce 
more than $35\%$ excess in $F_2$ because of the BCDMS data 
constraint.
In figure \ref{fqual2} the chosen modified parametrizations 
are shown together with the BCDMS data and the CTEQ4 parametrization.

\section{Conclusions}
This study shows (in agreement with 
previous estimates) that even largely anomalous and un\-con\-ventional
behaviour of the PDFs at high x, 
produces effects comparable
to statistical and systematic uncertainties, 
at the highest experimentally accessible $x$ and \( Q^{2} \)
regimes. It seems unlikely that any fine tuning of
the PDFs can produce a factor of two or higher 
excesses at high \( Q^{2} \).


\end{document}